\documentclass[useAMS,usenatbib]{mn2e}
\usepackage{latexsym}
\usepackage{makeidx}
\usepackage{cite}
\usepackage{amsmath}
\usepackage{color}
\usepackage{upgreek}
\usepackage{morefloats}
\usepackage{lineno}
\usepackage{setspace}
\usepackage{graphicx}

\title[Habitable worlds with JWST]{Habitable worlds with JWST: transit spectroscopy of the TRAPPIST-1 system?}
\author[J.~K. Barstow \& P.~.G.~.J. Irwin]{J.~K. Barstow$^{1,2}$\thanks{E-mail:
joannakbarstow@cantab.net (JKB)},
P.~G.~J. Irwin$^{2}$\\
$^{1}$Physics and Astronomy, University College London, London, UK\\
$^{2}$Atmospheric, Oceanic and Planetary Physics, Clarendon Laboratory, Department of Physics, University of Oxford, UK\\}

\begin{document}

\date{Submitted *** 2016}

\pagerange{\pageref{firstpage}--\pageref{lastpage}} \pubyear{2016}

\maketitle

\label{firstpage}

\begin{abstract}
The recent discovery of three Earth-sized, potentially habitable planets around a nearby cool star, TRAPPIST-1, has provided three key targets for the upcoming \textit{James Webb Space Telescope (JWST)}. Depending on their atmospheric characteristics and precise orbit configurations, it is possible that any of the three planets may be in the liquid water habitable zone, meaning that they may be capable of supporting life. We find that present-day Earth levels of ozone, if present, would be detectable if \textit{JWST} observes 60 transits for innermost planet 1b and 30 transits for 1c and 1d.  
\end{abstract}


\maketitle

\section{Introduction}
The infrared \textit{James Webb Space Telescope} (\textit{JWST}), due to launch in 2018, is predicted to dramatically change our understanding of exoplanet atmospheres. Included in this is the tantalising possibility that, if a suitable target is obtained, \textit{JWST} might provide the first atmospheric data for an Earth-sized planet orbiting in the habitable zone of its parent star. 

The \textit{TRAnsiting Planets and PlanetesImals Small Telescope} (\textit{TRAPPIST}, \citealt{jehin11}) is a 60-cm robotic telescope at La Silla observatory. It was designed for detection and characterisation of exoplanets, as well as observations of small solar system bodies. The recent discovery of the TRAPPIST-1 planetary system \citep{gillon16} has provided not one, but three, potential targets for \textit{JWST} follow up. TRAPPIST-1 is an ultracool dwarf of spectral type M8, only 12 parsecs away and hosting three planets with $R<1.2R_{\oplus}$. The innermost two planets b and c have 4$\times$ and 2$\times$ the irradiation experienced by Earth; the orbital period of the outermost planet d is not yet constrained, but the most likely period of 18.2 days would make it slightly cooler than Earth.
\begin{table*} \centering \begin{tabular}[c]{|c|c|c|c|c|c|c|} 
\hline
Planet & R (R$_{\oplus}$) & M (M$_{\oplus}$ - est.) & g (ms$^{-2}$ - est.) & $T_{\mathrm{eq}}$ (K) & H (km - est.) & Period (days)\\
\hline
1b & 1.113 $\pm$ 0.044 & 1.38 & 11 & 285---400 & 7.4---10.4& 1.510848 $\pm$ 0.000019\\  
1c & 1.049 $\pm$ 0.050 & 1.15 & 10 & 242---342 &  6.7---9.4 & 2.421848 $\pm$ 0.000028\\
1d & 1.168 $\pm$ 0.068 & 1.60 & 11 & 75---280 & 1.8---6.9 & 4.551---72.820 (18.202 most likely)\\
\end{tabular}
\caption{Basic parameters for the newly-discovered TRAPPIST-1 planets, taken from \citet{gillon16}. The orbit of TRAPPIST-1d is not yet constrained. Masses, scale heights and gravities are estimates based on the measured radii and T$_{\mathrm{eq}}$ range,  and the assumption that each planet has a bulk density equal to Earth's. Scale height and gravity are quoted at 1 bar.
\label{system}} 
\end{table*}

Depending on their albedos and the presence or absence of strong greenhouse warming, all of these planets could have the appropriate conditions for liquid water to be present. As this is considered to be a likely requirement for the presence of (Earth-like) life, this is an important criterion for any of these planets to be inhabited. \citet{bolmont16} have modelled the escape of H$_2$O during the evolution of each of these planets and find that, whilst there is a strong possibility that TRAPPIST-1b and 1c may have lost substantial amounts of water during their early life, cooler planet TRAPPIST-1d may have retained a substantial amount provided it started off with a sufficiently high water content. \citet{wheatley16} have observed the star in the X-ray with \textit{XMM-NEWTON} and find that the likely combined X-ray and EUV budget could be 50$\times$ higher than that assumed by \citet{bolmont16}, which increases the likelihood that 1b and 1c are dry planets. The likely chemistry and observational possibilities for habitable or inhabited worlds evolving around M-dwarf stars has been discussed at length by various authors, including \citet{segura05,deming09,kaltenegger09,rugheimer15} and \citet{tabataba15}. Predicted abundances of possible biosignature gases such as O$_3$ are highly dependent on a variety of factors, including the level of stellar activity and the UV flux profile of the star. This is still only known for a handful of M dwarfs and information is even scarcer for ultracool dwarfs, making it difficult to determine whether substantial amounts of O$_3$ could be sustained on a terrestrial planet orbiting a star like TRAPPIST-1. 
\begin{figure}
\centering
\includegraphics[width=0.45\textwidth]{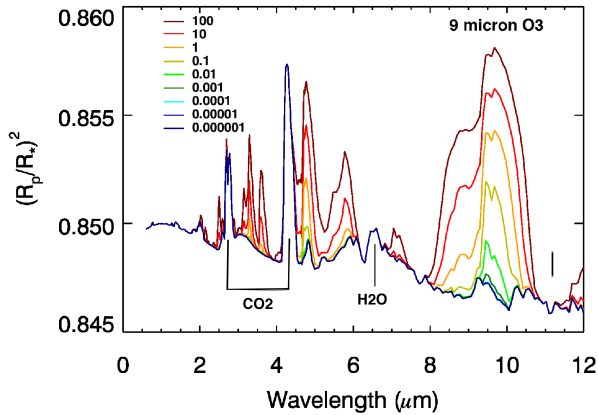}
\caption{Synthetic spectra assuming bulk properties of TRAPPIST-1d with an equilibrium temperature of 280 K, for different stratospheric abundances of O$_3$. This provides the largest transit signal out of the three planets. Stepping down from maroon through to navy, O$_3$ abundances go from 100$\times$ the present day Earth profile to 10$^{-6}\times$. Present day Earth is shown in orange. The black bar to the right of the plot indicates the 1$\sigma$ error bar for TRAPPIST-1d in the region of the 9 $\upmu$m O$_3$ feature, assuming 90 transits are observed with the \textit{JWST}/MIRI instrument. This is the smallest error bar obtained. The 9-$\upmu$m O$_3$ feature for an abundance of 10$^{-2}\times$ the Earth value is above the noise floor for this case, but O$_3$ at lower abundances would appear to be undetectable.\label{ozone}}
\end{figure}
In this paper we aim to see whether ozone might be detected by JWST in the atmospheres of the TRAPPIST-1 planets and make the following assumptions: 1) each TRAPPIST planet is capable of retaining liquid water, and therefore of hosting life; 2) on each planet, life has evolved and resulted in an atmosphere with $\sim$ 20\% molecular oxygen; 3) on each planet, a stratospheric ozone layer has formed, with ozone column abundance comparable to that of the Earth. For a detailed discussion of assumptions about ozone chemistry we refer the reader to \citet{barstow16} and references therein, in which we consider the case of an Earth-like planet orbiting an M5 star at 10 pc. The assumption that liquid water is retained is most likely to be valid for TRAPPIST-1d. 
\section{The TRAPPIST-1 system}
The TRAPPIST-1 system consists of three planets (b, c, d from the star outwards), of which the orbits of b and c are constrained. The orbital period of planet d is not yet determined, so a wide range of temperatures are possible for this planet. The most likely period is determined to be 18.202 days, which is the value we retain when calculating noise models. A summary of the system parameters is given in Table~\ref{system}. The system is located very close to the celestial equator \citep{skrutskie06}, so unfortunately is not within the polar continuous viewing zone of \textit{JWST}. This means that the maximum continuous visibility duration will only be of order 50 days, with long gaps where the system cannot be observed and observations only possible for around 100 days per year. 
\begin{figure*}
\centering
\includegraphics[width=0.9\textwidth]{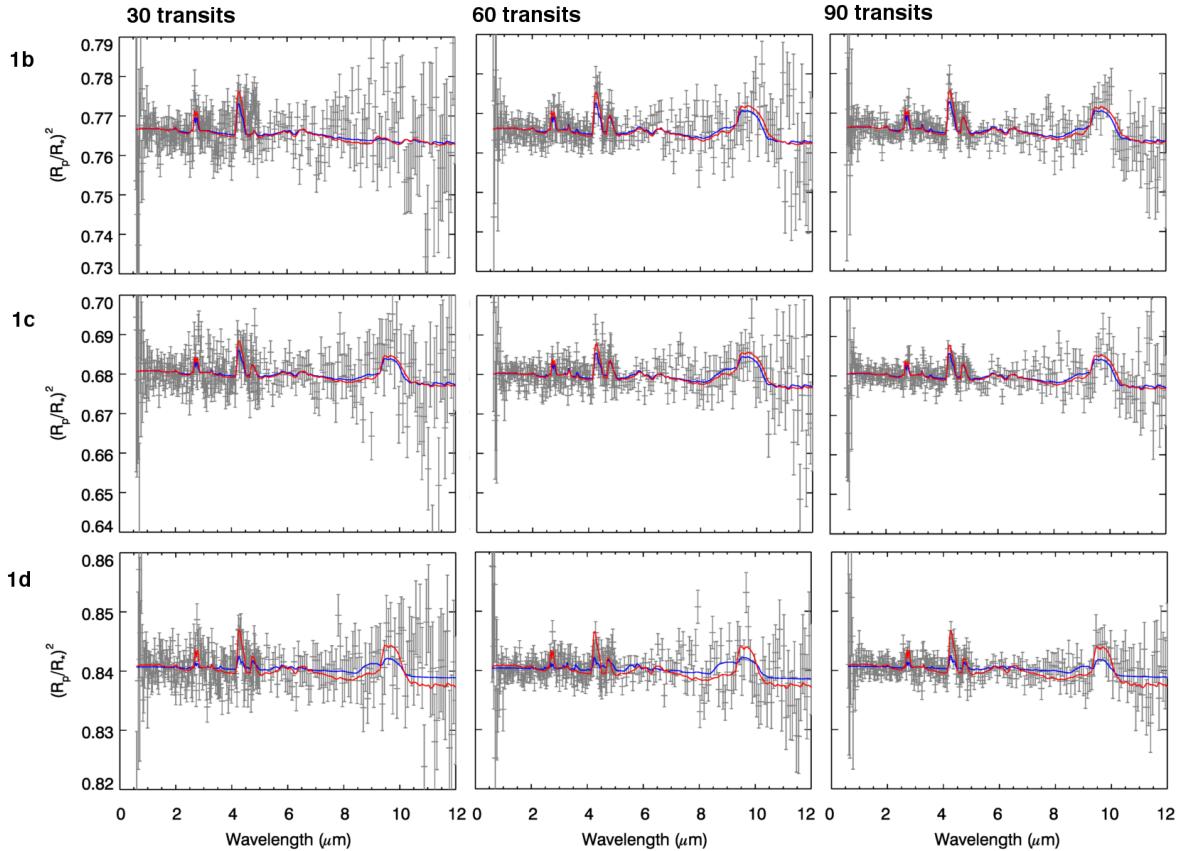}
\caption{Simulated \textit{JWST} observations of the TRAPPIST-1 planets, assuming 30, 60 and 90 transits are observed. Fits to the synthetic observations are shown for each case in blue (coldest equilibrium temperature) and red (hottest equilibrium temperature). For TRAPPIST 1b, at least 60 transits would be required with each instrument for O$_3$ to be detected, but for 1c and 1d 30 is sufficient.\label{spectra}}
\end{figure*}
\section{Model atmospheres}
We use the \textit{NEMESIS} radiative transfer and retrieval code \citep{irwin08} to simulate transmission spectra of the TRAPPIST-1 planets, under the assumption that each of them could have an Earth-like atmosphere. \textit{NEMESIS} couples a fast correlated-k \citep{goodyyung,lacis91} radiative transfer model with an optimal estimation retrieval algorithm \citep{rodg00}. It has been extensively used to model both exoplanets and solar system worlds (e.g. \citealt{lee12,tsang10,fletcher11}). 

Each planet is treated as though the atmosphere at the terminator, the region probed in transmission spectroscopy, has identical chemistry to Earth's present day atmosphere. The likelihood of this scenario for planets around cool stars is discussed in detail by \citet{barstow16} and references therein. The temperature profile is shifted from the present-day Earth case according to the assumed equilibrium temerature of each planet; we take this to be the mean temperature from the possible range indicated by the observations, corresponding to 343 K, 292 K and 180 K respectively for planets b, c and d. For comparison, Earth's equilibrium temperature is 255 K. The Earth atmospheric model is based on that used by \citet{irwin14} and later by \citet{barstow15,barstow16}. Gas absorption data are taken from the HITRAN08 database \citep{rothman09}. H$_2$O clouds are also included in the model, although since they are relatively deep in the atmosphere they do not have a significant effect on the spectrum. No masses have yet been measured for these planets, but we estimate masses assuming that all planets have the same bulk density as the Earth (Table~\ref{system}).

The stellar spectrum used is taken from the PHOENIX model atmosphere library \citep{husser13}, for a 2500 K, solar metallicity star with a log($g$) of 5.0. We choose the cooler available model of 2500 K over 2600 K, although the star temperature is 2550 K, because a fainter star provides a more conservative error estimate. The spectrum is extrapolated as a black body at wavelengths longer than 5 $\upmu$m, for which no model information is available. The stellar radius of 81373.5 km is taken from \citep{gillon16}. 

Whilst a surface pressure consistent with Earth's present day atmosphere is used for all input models, for the retrieval we extend the atmosphere to 90 bar, the surface pressure of Venus, since we have no way of telling from initial observations whether the planet has a surface at 1 bar or at higher (or indeed lower) pressures. The radius at the bottom of the atmosphere is a variable in the retrieval, which tests our ability to account for lack of knowledge of the surface pressure. For each planet we perform two retrievals assuming fixed isothermal temperature profiles, at stratospheric temperatures based on the maximum and minimum equlibrium temperatures as given in Table~\ref{system}, where $T_{strat}=T_{eq}/2^{0.25}$. 

Most importantly, the \textit{a priori} model atmosphere in the retrieval contains a very low (undetectable) abundance of ozone. The retrievals we perform therefore test whether an ozone signature could be detected from the spectra. An approximately Earth-like abundance of ozone will only appear in the retrieved atmospheric state vector if there is information in the spectrum that suggests it should be there. In Figure~\ref{ozone} we show the effect on the transmission spectrum of changing the O$_3$ abundance. As the abundance decreases below the present day Earth value the feature disappears rapidly and becomes undetectable. Anything below 10$^{-2}\times$ the Earth present day O$_3$ value would appear not to be observable.
\section{JWST and noise model}
We model the planets assuming that they will be observed by both the Near InfraRed Spectrograph (NIRSpec) and Mid InfraRed Instrument (MIRI) Low Resolution Spectrometer (LRS) on \textit{JWST}. NIRSpec is assumed to be used with the prism, providing a continuous spectrum from 0.6 to 5 $\upmu$m. The noise model used for \textit{JWST} is adapted from \citet{barstow15,barstow16}. The photon noise is calculated using the following equation:
\begin{equation}
n_{\lambda}=I_{\lambda}{\pi}(r_{\star}/D_{\star})^2({\lambda}/hc)({\lambda}/R)A_{eff}Q{\eta}t
\end{equation}
where $n_{\lambda}$ is the number of photons received for a given wavelength $\lambda$, $I_{\lambda}$ is the spectral radiance of the stellar signal, $r_{\star}$ is the stellar radius, $D_{\star}$ is the distance to the star, $h$ and $c$ are the Planck constant and speed of light, $R$ is the spectral resolving power, $A_{eff}$ is the telescope effective area, $Q$ is the detector quantum efficiency, $\eta$ is the the throughput and $t$ is the exposure time. The effective exposure time is taken to be the transit duration from Table~\ref{system}, assuming an 80\% duty cycle, and therefore planets with shorter transit durations will have noisier spectra. Values for all instrument parameters are identical to those used by \citet{barstow15,barstow16}, and system specific values are as discussed elsewhere in this work. Noise added to the synthetic spectra is random and white - no correlated noise is included. To calculate the noise on the stellar radiance $I$ we take the square root of the number of photons and then invert the above equation. The noise on the transit depth is given by $\sqrt{2}\times\sigma_{I}/I$. 
\section{Results}
\begin{figure}
\centering
\includegraphics[width=0.45\textwidth]{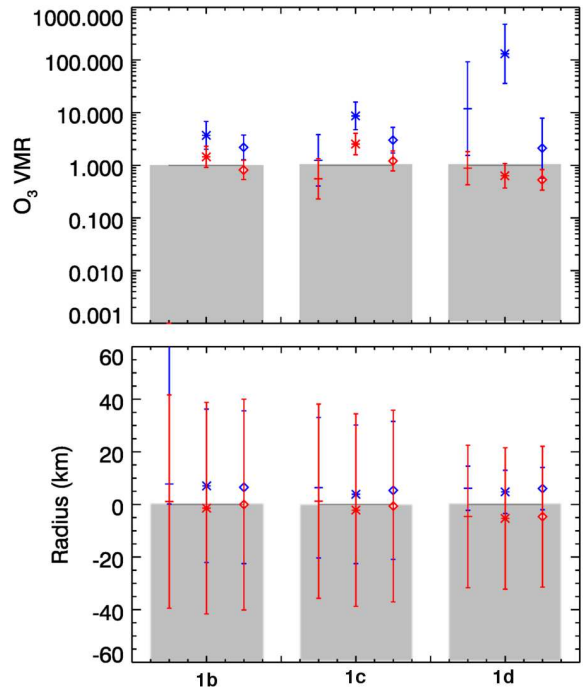}
\caption{Retrieval results for O$_3$ abundance and 1-bar radius. Input values are indicated by grey bars. Retrieved values are coloured blue and red for cold/hot temperature retrievals. Crosses/asterisks/diamonds correspond to 30/60/90 transits. No 30 transit points are shown in the top panel for 1b as no O$_3$ detection was possible. There is degeneracy between the retrieved parameters, with smaller/larger retrieved radii compensating for hotter/colder temperatures; hotter atmospheres have a larger scale height, as do planets with a larger radius for a given mass. The balance of this degeneracy affects the retrieved O$_3$ abundance, as seen for the cold TRAPPIST-1d scenario with 60 transits, for which O$_3$ is significantly over-retrieved to compensate for a small atmospheric scale height.\label{retrieved}}
\end{figure}

 We present simulated spectra assuming 30, 60 and 90 transits of each instrument (NIRSpec and MIRI), with spectral fits from retrievals based on the extreme equilibrium temperatures for each planet (coldest blue, hottest red; Figure~\ref{spectra}). Retrieved properties are the O$_3$ volume mixing ratio (VMR), planetary radius at the solid surface (90 bar pressure level), and H$_2$O and CO$_2$ VMRs. The H$_2$O and CO$_2$ VMRs do not deviate far from the prior and we conclude that the retrieval is relatively insensitive to these properties. We present the retrieval results for radius offset and O$_3$ VMR in Figure~\ref{retrieved}. Here, the radius offset quoted is at 1 bar to facilitate easy comparison with the true value. The retrieved radius compensates for temperature deviations from the true value, as it is smaller for the higher temperature retrievals and larger for the lower temperature ones. As well as just increasing the size of the planet, increasing the radius also increases the scale height as the gravity is slightly reduced. 

The \textit{a priori} abundance of O$_3$ is set to be 10$^{-8}\times$ the present-day Earth value. This value is low enough such that no O$_3$ features are visible at all in the spectrum. We then retrieve a scaling factor on the present day Earth profile starting from this prior assumption. Our results show that O$_3$, if present in quantities similar to present-day Earth, would be detectable on all TRAPPIST-1 planets if at least 60 transits are obtained with both NIRSpec and MIRI. 

TRAPPIST-1d is the most likely of the three planets to be Earthlike. O$_3$ can be detected for 30 transits each of NIRSpec and MIRI, regardless of the temperature profile used in the retrieval. CO$_2$ features can also be seen at shorter wavelengths, with the 4.3 $\upmu$m feature visible above the noise in most cases. Where detected, the O$_3$ feature looks very different for the cold and hot TRAPPIST-1d temperature profiles. This is because, to compensate for the small scale height for the cooler 80K profile, the retrieved O$_3$ abundance is more than an order of magnitude higher than the input value for all noise levels. This alteration in band shape for high O$_3$ abundances can clearly be seen comparing Figures~\ref{ozone} and ~\ref{spectra}. This is because the scale height is smaller for cooler temperatures, so the retrieval compensates by increasing the O$_3$ abundance. Conversely, the O$_3$ abundance is slightly under-retrieved for the hotter temperature retrievals. Therefore, a reliable detection of the absolute O$_3$ abundance would be challenging for these planets.
\section{Discussion}
\label{discussion}
We find that O$_3$ at present-day Earth levels would be detectable for TRAPPIST-1c and -1d if at least 30 transits each with NIRSpec and MIRI are observed, and for TRAPPIST-1b with 60 transits. However, TRAPPIST-1b and TRAPPIST-1c are likely to be hotter than present-day Earth, and may in fact have very different atmospheres. The fact that we could detect O$_3$ (and CO$_2$ features are also clearly visible) indicates that these would be interesting targets regardless of their atmospheric chemistry as other molecular species might be similarly detectable. 

Regarding TRAPPIST-1d it is likely that, even if life evolves and produces an atmosphere with approximately 20\% O$_2$ and broadly Earth-like conditions, the photochemical processes producing O$_3$ will be very different. The extent to which this is true is difficult to determine at present, due to a lack of detailed information about the host star. UV and further X-ray observations of the host's output could provide constraints for photochemical models, that might allow a prediction of the likely O$_3$ abundance. The possibility remains that, even in the case where the atmosphere is oxygen rich, insufficient O$_3$ will be produced for detection.  Detailed chemical modelling of all three planets will be necessary prior to \textit{JWST} observations to determine whether observable signatures are likely, as this will be an important consideration for time allocation commitees. 
Approximately 30 transits with each instrument are needed to detect O$_3$ on TRAPPIST-1d, so 60 transits in total. Since the TRAPPIST system is located close to the celestial equator, and TRAPPIST-1d has a likely orbital period of $\sim$18 days, this may present a problem. If TRAPPIST-1 is visible for around 100 days per year, then only 5 or 6 transits per year could be observed. Assuming JWST remains operational for 10 years, this would just allow us to reach the 60 transit mark. If biosignature detection is the goal, the obvious choice for the sake of economy would be to observe with MIRI only, as the 9-$\upmu$m O$_3$ feature is the most prominent. We performed this test, and find that detecting O$_3$ may be possible with 30 transits of MIRI alone, but the dependence on the temperature used in the retrieval becomes more critical. The abundance is under retrieved for the high temperature case and over retrieved for the low temperature case to a greater extent. A possible alternative to obtaining 30 transits with each of NIRSpec and MIRI would be to obtain photometry over a handful of transits at shorter wavelengths, which could help to break these degeneracies. 

TRAPPIST-1d may have a closer and shorter orbit. If, for example, the orbit is 10 days, the problem becomes more tractable, with 30 transits each with NIRSpec and MIRI observable within 6 years. The transit duration will decrease slightly, and this will increase the noise level on the spectrum, but the signal to noise will still be better than for 1b and 1c. Of course, the orbits for TRAPPIST-1b and 1c are considerably shorter, with periods of only 1.5 and 2.4 days. Although these planets are probably less likely to be Earth-like due to their hotter temperatures, 60 or even 90 transits with each instrument would be far more easily accomplished. 180 transits of TRAPPIST-1b could be accomplished in 270 days, so within 3 years of JWST observations.

Regardless of the adopted strategy, \textit{JWST} observations of this new system are likely to be a significant challenge, and final results may be the result of several years of accumulated data. However, it is hoped that this discovery is only the first of many similar systems. The TRAPPIST survey is a prototype for a much more ambitious project, the \textit{Search for habitable Planets EClipsing ULtra-cOOl Stars} (\textit{SPECULOOS} \citealt{gillon13}). Targeting further ultracool stars may result in the discovery of similar systems to TRAPPIST-1, but closer by, or within the \textit{JWST} continuous viewing zone. Such a system within 10 pc and closer to the celestial pole would present exciting opportunities. 
\section{Conclusions}
We perform radiative transfer simulations for the newly discovered TRAPPIST-1 planets, under the assumption that each may have an atmosphere similar to that of present-day Earth. We find that the biosignature ozone, at present-day Earth levels, could be detectable for TRAPPIST-1b with at least 60 transits each with NIRSpec and MIRI, and for 1c and 1b with 30 transits. TRAPPIST-1d, as the coolest planet in the system, is most likely to be habitable and therefore have substantial amounts of ozone; however, this planet is currently thought to have an 18-day orbit, and the system is far from the continuous viewing zone for \textit{JWST}, meaning that 30 transits with each instrument would take 10 years to obtain. The inner planets 1b and 1c are more favourable targets from this perspective, due to their much shorter orbits, although they are probably too warm and heavily irradiated to have Earth-like atmospheres.  

As a first step in preparation, detailed photochemical modelling of this system is necessary, based on further measurements of the star in the X-ray and UV if possible. This will allow us to determine the likelihood of surface liquid water, and of biosignature gases being present, on all three planets. Determination of radial velocity masses, whilst likely to be challenging, would also be highly desirable. This first planetary system around a nearby ultracool dwarf is very promising, and it is to be hoped that further systems will be discovered in future with both \textit{TRAPPIST} and \textit{SPECULOOS}.
\section*{Acknowledgements}
JKB is funded under the ERC project 617119 (ExoLights) and PGJI acknowledges the support of the Science and Technology Facilities Council. We thank Adam Burgasser for a helpful discussion and the anonymous reviewer for their comments. 
\bibliographystyle{mn2e}
\bibliography{bibliography}

\begin{thebibliography}{}

\bibitem[\protect\citeauthoryear{{Barstow}, {Aigrain}, {Irwin}, {Kendrew} \&
  {Fletcher}}{{Barstow} et~al.}{2015}]{barstow15}
{Barstow} J.~K.,  {Aigrain} S.,  {Irwin} P.~G.~J.,  {Kendrew} S.,    {Fletcher}
  L.~N.,  2015, MNRAS, 448, 2546

\bibitem[\protect\citeauthoryear{{Barstow}, {Aigrain}, {Irwin}, {Kendrew} \&
  {Fletcher}}{{Barstow} et~al.}{2016}]{barstow16}
{Barstow} J.~K.,  {Aigrain} S.,  {Irwin} P.~G.~J.,  {Kendrew} S.,    {Fletcher}
  L.~N.,  2016, MNRAS, 458, 2657

\bibitem[\protect\citeauthoryear{{Bolmont}, {Selsis}, {Owen}, {Ribas},
  {Raymond}, {Leconte} \& {Gillon}}{{Bolmont} et~al.}{2016}]{bolmont16}
{Bolmont} E.,  {Selsis} F.,  {Owen} J.~E.,  {Ribas} I.,  {Raymond} S.~N.,
  {Leconte} J.,    {Gillon} M.,  2016, ArXiv e-prints

\bibitem[\protect\citeauthoryear{{Deming}, {Seager}, {Winn}, {Miller-Ricci},
  {Clampin}, {Lindler}, {Greene}, {Charbonneau}, {Laughlin}, {Ricker}, {Latham}
  \& {Ennico}}{{Deming} et~al.}{2009}]{deming09}
{Deming} D.,  {Seager} S.,  {Winn} J.,  {Miller-Ricci} E.,  {Clampin} M.,
  {Lindler} D.,  {Greene} T.,  {Charbonneau} D.,  {Laughlin} G.,  {Ricker} G.,
  {Latham} D.,    {Ennico} K.,  2009, {PASP}, 121, 952

\bibitem[\protect\citeauthoryear{{Fletcher}}{{Fletcher}}{2011}]{fletcher11}
{Fletcher} L.~N. e.~a.,  2011, Science, 332, 1413

\bibitem[\protect\citeauthoryear{{Gillon}, {Jehin}, {Delrez}, {Magain},
  {Opitom} \& {Sohy}}{{Gillon} et~al.}{2013}]{gillon13}
{Gillon} M.,  {Jehin} E.,  {Delrez} L.,  {Magain} P.,  {Opitom} C.,    {Sohy}
  S.,  2013, in Protostars and Planets VI Posters {SPECULOOS: Search for
  habitable Planets EClipsing ULtra-cOOl Stars}.
p.~66

\bibitem[\protect\citeauthoryear{{Gillon}, {Jehin} \& {Lederer}}{{Gillon}
  et~al.}{2016}]{gillon16}
{Gillon} M.,  {Jehin} E.,    {Lederer} S.~M. e.~a.,  2016, Nature

\bibitem[\protect\citeauthoryear{{Goody} \& {Yung}}{{Goody} \&
  {Yung}}{1989}]{goodyyung}
{Goody} R.~M.,  {Yung} Y.~L.,  1989, {Atmospheric radiation : theoretical
  basis}

\bibitem[\protect\citeauthoryear{{Husser}, {Wende-von Berg}, {Dreizler},
  {Homeier}, {Reiners}, {Barman} \& {Hauschildt}}{{Husser}
  et~al.}{2013}]{husser13}
{Husser} T.-O.,  {Wende-von Berg} S.,  {Dreizler} S.,  {Homeier} D.,  {Reiners}
  A.,  {Barman} T.,    {Hauschildt} P.~H.,  2013, A\&A, 553, A6

\bibitem[\protect\citeauthoryear{{Irwin}, {Barstow}, {Bowles}, {Fletcher},
  {Aigrain} \& {Lee}}{{Irwin} et~al.}{2014}]{irwin14}
{Irwin} P.~G.~J.,  {Barstow} J.~K.,  {Bowles} N.~E.,  {Fletcher} L.~N.,
  {Aigrain} S.,    {Lee} J.-M.,  2014, Icarus, 242, 172

\bibitem[\protect\citeauthoryear{{Irwin}, {Teanby}, {de Kok}, {Fletcher},
  {Howett}, {Tsang}, {Wilson}, {Calcutt}, {Nixon} \& {Parrish}}{{Irwin}
  et~al.}{2008}]{irwin08}
{Irwin} P.~G.~J.,  {Teanby} N.~A.,  {de Kok} R.,  {Fletcher} L.~N.,  {Howett}
  C.~J.~A.,  {Tsang} C.~C.~C.,  {Wilson} C.~F.,  {Calcutt} S.~B.,  {Nixon}
  C.~A.,    {Parrish} P.~D.,  2008, Journal of Quantitative Spectorscopy and
  Radiative Transfer, 109, 1136

\bibitem[\protect\citeauthoryear{{Jehin}, {Gillon}, {Queloz}, {Magain},
  {Manfroid}, {Chantry}, {Lendl}, {Hutsem{\'e}kers} \& {Udry}}{{Jehin}
  et~al.}{2011}]{jehin11}
{Jehin} E.,  {Gillon} M.,  {Queloz} D.,  {Magain} P.,  {Manfroid} J.,
  {Chantry} V.,  {Lendl} M.,  {Hutsem{\'e}kers} D.,    {Udry} S.,  2011, The
  Messenger, 145, 2

\bibitem[\protect\citeauthoryear{{Kaltenegger} \& {Traub}}{{Kaltenegger} \&
  {Traub}}{2009}]{kaltenegger09}
{Kaltenegger} L.,  {Traub} W.~A.,  2009, The Astrophysical Journal, 698, 519

\bibitem[\protect\citeauthoryear{{Lacis} \& {Oinas}}{{Lacis} \&
  {Oinas}}{1991}]{lacis91}
{Lacis} A.~A.,  {Oinas} V.,  1991, J. Geophys. Res., 96, 9027

\bibitem[\protect\citeauthoryear{{Lee}, {Fletcher} \& {Irwin}}{{Lee}
  et~al.}{2012}]{lee12}
{Lee} J.-M.,  {Fletcher} L.~N.,    {Irwin} P.~G.~J.,  2012, Monthly Notices of
  the Royal Astronomical Society, 420, 170

\bibitem[\protect\citeauthoryear{{Rodgers}}{{Rodgers}}{2000}]{rodg00}
{Rodgers} C.~D.,  2000, Inverse Methods for Atmospheric Sounding.
{World Scientific}

\bibitem[\protect\citeauthoryear{{Rothman}}{{Rothman}}{2009}]{rothman09}
{Rothman} L.~S. e.~a.,  2009, Journal of Quantitative Spectroscopy and
  Radiative Transfer, 110, 533

\bibitem[\protect\citeauthoryear{{Rugheimer}, {Kaltenegger}, {Segura}, {Linsky}
  \& {Mohanty}}{{Rugheimer} et~al.}{2015}]{rugheimer15}
{Rugheimer} S.,  {Kaltenegger} L.,  {Segura} A.,  {Linsky} J.,    {Mohanty} S.,
   2015, The Astrophysical Journal, 809, 57

\bibitem[\protect\citeauthoryear{{Segura}, {Kasting}, {Meadows}, {Cohen},
  {Scalo}, {Crisp}, {Butler} \& {Tinetti}}{{Segura} et~al.}{2005}]{segura05}
{Segura} A.,  {Kasting} J.~F.,  {Meadows} V.,  {Cohen} M.,  {Scalo} J.,
  {Crisp} D.,  {Butler} R.~A.~H.,    {Tinetti} G.,  2005, Astrobiology, 5, 706

\bibitem[\protect\citeauthoryear{{Skrutskie}, {Cutri} \&
  {Stiening}}{{Skrutskie} et~al.}{2006}]{skrutskie06}
{Skrutskie} M.~F.,  {Cutri} R.~M.,    {Stiening} R. e.~a.,  2006, AJ, 131, 1163

\bibitem[\protect\citeauthoryear{{Tabataba-Vakili}, {Grenfell},
  {Grie{\ss}meier} \& {Rauer}}{{Tabataba-Vakili} et~al.}{2015}]{tabataba15}
{Tabataba-Vakili} F.,  {Grenfell} J.~L.,  {Grie{\ss}meier} J.-M.,    {Rauer}
  H.,  2015, ArXiv e-prints

\bibitem[\protect\citeauthoryear{{Tsang}, {Wilson}, {Barstow}, {Irwin},
  {Taylor}, {McGouldrick}, {Piccioni}, {Drossart} \& {Svedhem}}{{Tsang}
  et~al.}{2010}]{tsang10}
{Tsang} C.~C.~C.,  {Wilson} C.~F.,  {Barstow} J.~K.,  {Irwin} P.~G.~J.,
  {Taylor} F.~W.,  {McGouldrick} K.,  {Piccioni} G.,  {Drossart} P.,
  {Svedhem} H.,  2010, Geophysical Research Letters, 37, 2202

\bibitem[\protect\citeauthoryear{{Wheatley}, {Louden}, {Bourrier}, {Ehrenreich}
  \& {Gillon}}{{Wheatley} et~al.}{2016}]{wheatley16}
{Wheatley} P.~J.,  {Louden} T.,  {Bourrier} V.,  {Ehrenreich} D.,    {Gillon}
  M.,  2016, ArXiv e-prints

\end{thebibliography}

\label{lastpage}

\end{document}